\documentclass[aps, pra, twocolumn, superscriptaddress]{revtex4-1}
\usepackage{graphicx}
\usepackage{amssymb}
\usepackage{amsmath}
\usepackage{booktabs}
\usepackage[colorlinks=True,linkcolor=blue,anchorcolor=blue,citecolor=blue,CJKbookmarks=true]{hyperref}
\usepackage{bm}
\usepackage{makecell}

\usepackage{ulem}

\begin{document}
\title{Electron-positron pair creation in a supercritical static asymmetric potential well}

\author{Z. L. Li}
\email{zlli@cumtb.edu.cn}
\affiliation{School of Science, China University of Mining and Technology, Beijing 100083, China}

\author{A. R. Sun}
\affiliation{School of Science, China University of Mining and Technology, Beijing 100083, China}

\author{J. H. Xia}
\affiliation{School of Science, China University of Mining and Technology, Beijing 100083, China}

\author{J. X. Wu}
\affiliation{School of Science, China University of Mining and Technology, Beijing 100083, China}

\author{Y. J. Li}
\email{lyj@aphy.iphy.ac.cn}
\affiliation{School of Science, China University of Mining and Technology, Beijing 100083, China}
\affiliation{State Key Laboratory for Tunnel Engineering, China University of Mining and Technology, Beijing 100083, China}

\date{\today}

\begin{abstract}
The electron-positron pair creation in a supercritical static asymmetric potential well, which is composed of a subcritical and a supercritical potential separated by a fixed distance, is investigated using computational quantum field theory. To explain the discrete peaks in the positron energy spectrum, an analytical formula for determining the positions of bound states in a subcritical asymmetric potential well is derived and extended to the supercritical asymmetric potential well in two ways. One of the two methods can not only predict the positions of bound states, but also offer the pair creation rate. This study also reveals that the subcritical potential height can optimize the energy spread of created electrons, providing a new way to produce high-energy electron beams with concentrated energy in experiments. Moreover, it is found that the pair creation rate in a supercritical asymmetric potential well, composed of a subcritical symmetric potential well and a supercritical Sauter potential, exceeds the sum of the pair creation rates produced by each potential individually. This finding suggests a potential method for enhancing pair yield.
\end{abstract}

\maketitle

\section{INTRODUCTION}\label{sec:one}

A sufficiently strong external field can breakdown the vacuum, leading to the spontaneous creation of electron-positron pairs \cite{Dirac19301,Sauter1931,Heisenberg1936,Schwinger1951}. This phenomenon has become one of the most striking predictions of quantum electrodynamics (QED). The electric field strength required to breakdown the vacuum is equal to $E_{\mathrm{cr}}=m^2c^3/(e\hbar)\sim10^{16}\mathrm{V}/\mathrm{cm}$, where $m$ and $-e$ are the mass and charge of electrons, respectively. This field strength is also called Schwinger critical field strength. Due to the non-perturbative nature of Schwinger pair production, it also has profound implications for understanding other non-perturbative processes under extreme conditions, such as astrophysical environments and high-intensity laser experiments \cite{Xie2017,Fedotov2023}.
Despite its theoretical elegance, the experimental observation of Schwinger pair production remains challenging, owing to the requirement for extremely high field strengths. However, recent advances in laser technology and the development of ultra-intense laser facilities have brought us closer to realizing these conditions experimentally \cite{ELI,XCELS}.
Given that experimental verification of vacuum pair production still has some distance to go, theoretical physicists are devoted to research in two areas. One is to explore new mechanisms for enhancing pair creation or reducing the threshold of pair production \cite{Schutzhold2008,Piazza2009,Bulanov2010,Titov2012,
Li2014,Torgrimsson2016,Torgrimsson2019,Xie2022}. The second is to discover new physical phenomena in pair production, achieving a more comprehensive and in-depth understanding of the pair production process \cite{Dumlu2010,Hebenstreit2011,Akkermans2012,
Kohlfurst2014,Li2015,Li2017,Lv2018,Su2019,Kohlfurst2019}.

In Ref. \cite{Lv2018}, the authors studied electron-positron pair creation from the vacuum in a double-step potential composed of a subcritical and a supercritical Sauter potential separated by a considerable distance. They discovered a clear nonlocal signature in the pair creation process: The subcritical potential can control the positron energy spectrum produced by the supercritical potential, though the created positrons had never visited the subcritical potential. In our recent work \cite{Wu2023}, we found that when the two potentials form a supercritical asymmetric potential well, the symmetry of the electron and positron energy spectra produced by the supercritical potential is broken, and the positron energy spectrum exhibits obvious discrete peaks. We believed that it was due to the emergence of resonant states formed by the overlap between the bound states in the potential well and the negative energy continuum. However, in contrast to the supercritical symmetric well, where a transmission coefficient formula can be used to determine the positions of resonant states \cite{Greiner4}, the supercritical asymmetric well does not possess such an analytical expression. Therefore, it is difficult to rigorously confirm the conjecture mentioned above. The difficulty lies in how to determine the positions of resonant states.
In fact, resonant states are widely present in atomic, molecular, nuclear physics and in chemistry.
The complex scaling method \cite{Reinhardt1982,
Moiseyev1998,Ivanov2004,Alhaidari2007,Lv2013,Lv2014,Zhou2023} is a powerful tool for numerically calculating positions and widths of resonant states.
Nevertheless, whether there are analytical methods has not been fully explored.

In this paper, we derive an analytical formula for the bound state energy levels in a subcritical asymmetric potential well and extend it to the supercritical case in two ways. By comparing the analytical results with those obtained from CQFT, the validity of these methods is demonstrated.
Moreover, employing computational quantum field theory (CQFT), we also investigate the effect of the subcritical potential height on the electron and positron energy spectra produced by the supercritical potential, and the time evolution of created particles in the supercritical asymmetric potential well. The findings will not only deepen our understanding of pair creation in a supercritical asymmetric potential well but also provide ways to optimize pair creation.

This paper is organized as follows. In Sect. \ref{sec:two}, we briefly introduce the computational quantum field theory (CQFT) and the field model we used. In Sect. \ref{sec:three}, we explore the analytical formula for the bound state energy levels in a supercritical asymmetric potential well. The effect of the subcritical potential height on the energy spectra of created  particles and the time evolution of created particles in the supercritical asymmetric potential well are also investigated. Section \ref{sec:four} is the conclusion and outlook. The calculation of the energy and width of resonant states in a supercritical symmetric potential well is shown in Appendix \ref{appa}.

\section{Theoretical method and model}
\label{sec:two}

The starting point for computational quantum field theory (CQFT) is the Dirac equation satisfied by the Dirac field operator: $\mathrm{i}\partial \hat{\psi}/ \partial t =h\hat{\psi}$, where $h=c\bm{\alpha} \cdot \mathbf{p}+\beta mc^2-eV(\mathbf{x}, t)$ is the Hamiltonian, $\bm{\alpha}=(\alpha_1, \alpha_2, \alpha_3)$ and $\beta$ are $4\times4$ Dirac matrices, $\mathbf{p}=-\mathrm{i}\mathbf{\nabla}$ is the canonical momentum operator, $c$ is the speed of light, $-e$ and $m$ are the charge and mass of electrons, and $V(\mathbf{x}, t)$ is the scalar potential. For convenience, we will use the atomic units $\hbar=e=m=1$ below. Since our focus is on the problem of  pair creation in one-dimensional static asymmetric potential wells, the Dirac equation can be further simplified. The reduced Hamiltonian has the form
\begin{equation}\label{eqn:Hamiltonian1}
h(z)=c \sigma_1 p_z+\sigma_3 c^2-V(z),
\end{equation}
where $\sigma_1$ and $\sigma_3$ are Pauli matrices.

In quantum field theory, the Dirac field operator can be expanded in terms of different complete bases:
\begin{equation}\label{eqn:Fourier}
\begin{split}
\hat{\psi}(z,t)&=\sum_p \hat{b}_p(t)u_p(z)+\sum_n \hat{d}_n^{\dagger}(t)v_n(z)\\
&=\sum_p \hat{b}_pu_p(z,t)+\sum_n \hat{d}_n^{\dagger}v_n(z,t),
\end{split}
\end{equation}
where $u_p(z)$ and $v_n(z)$ are the positive and negative energy states of the field-free Hamiltonian, $u_p(z,t)$ and $v_n(z,t)$ are the positive and negative energy states obtained from the time evolution of $u_p(z)$ and $v_n(z)$ under the whole Hamiltonian (\ref{eqn:Hamiltonian1}).                                $\hat{b}_p(t)$ and $\hat{d}_n^{\dagger}(t)$ are the time-dependent annihilation and creation operators, $\hat{b}_p$ and $\hat{d}_n^{\dagger}$ are the time-independent annihilation and creation operators.
These operators satisfy the standard anticommutation relations.
Furthermore, the time-dependent and time-independent operators can be related by the Bogoliubov transformation:
\begin{equation}\label{eqn:Relationship}
\begin{split}
\hat{b}_p(t)&=\sum_{p'} \hat{b}_{p'}U_{pp'}(t)+\sum_n \hat{d}_n^{\dagger}U_{pn}(t), \\
\hat{d}_n^{\dagger}(t)&=\sum_p \hat{b}_pU_{np}(t)+\sum_n \hat{d}_n^{\dagger}U_{nn'}(t),
\end{split}
\end{equation}
where $U_{pp'}(t)=\int \mathrm{d}z\,u_p^\dagger(z)u_{p'}(z,t)$, $U_{pn}(t)=\int \mathrm{d}z\,u_p^\dagger(z)v_{n}(z,t)$, $U_{np}(t)=\int \mathrm{d}z\,v_n^\dagger(z)u_{p}(z,t)$, and $U_{nn'}(t)=\int \mathrm{d}z\,v_n^\dagger(z)v_{n'}(z,t)$ are the matrix elements of the time evolution operator $U(t)=T\exp(-\mathrm{i}h t)$ ($T$ is the time-ordered operator), which can be calculated through the split-operator technique \cite{Braun1999,Mocken2008}. The probability density of created electrons is defined as
\begin{equation}\label{eqn:ProDen}
\begin{split}
\rho(z,t)&=\langle\mathrm{vac}|\hat{\psi}_{e}^\dagger(z,t)
\hat{\psi}_{e}(z,t)|\mathrm{vac}\rangle \\
&=\sum_n\Big|\sum_pU_{pn}(t)u_p(z)\Big|^2,
\end{split}
\end{equation}
where $\hat{\psi}_{e}(z,t)=\sum_p b_p(t)u_p(z)$ denotes the electron portion of the field operator. Integrating the probability density over space, we obtain the number of created electrons
\begin{equation}\label{eqn:NumDen}
N(t)=\int\!\mathrm{d}z\,\rho(z,t)=\sum_p N(p,t),
\end{equation}
where $N(p,t)=\sum_n| U_{pn}(t)|^2$ is called the momentum distribution function in CQFT. The energy distribution function is
\begin{equation}\label{eqn:EnergySpectrum}
N(E,t)=\frac{L}{2\pi (\mathrm{d}E/\mathrm{d}p)}\sum_n| U_{pn}(t)|^2,
\end{equation}
where $L$ is the length of the numerical box and $\mathrm{d}E/\mathrm{d}p$ represents the group velocity of the particle wave packet \cite{Wu2023}.

In our model, the asymmetric potential well consisted of two Sauter potentials is
\begin{equation}\label{eqn:Potential}
-V(z)=\frac{V_1}{2}[1+\tanh(z/w)]-\frac{V_2}{2}[1+\tanh(z+d/w)],
\end{equation}
where $V_1$ and $V_2$ are the two potential height, $w$ denotes the potential width, and $d$ is the distance between the two potentials or the potential well width. Notice that when the potential height is larger than $2c^2$, the potential is called a supercritical potential; when it is smaller than $2c^2$, it is named a subcritical potential.

\section{Numerical results}
\label{sec:three}
In this section, we will explore the method to determine the bound state energy levels in an asymmetric potential well, examine the effect of the subcritical potential height on the energy spectra of created particles, and discuss the time evolution of created pairs in the asymmetric potential well.

\subsection{Determination of the bound state energy levels in an asymmetric potential well}
\label{sec:sub3a}
To explain the discrete peak structures observed in the energy spectrum of positrons created in supercritical asymmetric potential wells (see Fig. 2 in \cite{Wu2023}), we derive an analytical formula for the bound state energy levels in subcritical asymmetric potential wells and extend it to supercritical potential wells using two methods.
The first method is to fit the functional relationship between the bound state energy and the larger potential height $V_1$ based on this analytical formula, and then apply it to the case where $V_1$ is supercritical.
The second one is to extend this analytical formula to the complex energy plane to compute the bound state energy levels in supercritical potential wells.
Furthermore, the feasibility of the two methods is demonstrated by comparing their computational results with those obtained from CQFT.

\begin{figure}[!ht]
\centering
\includegraphics[width=0.42\textwidth]{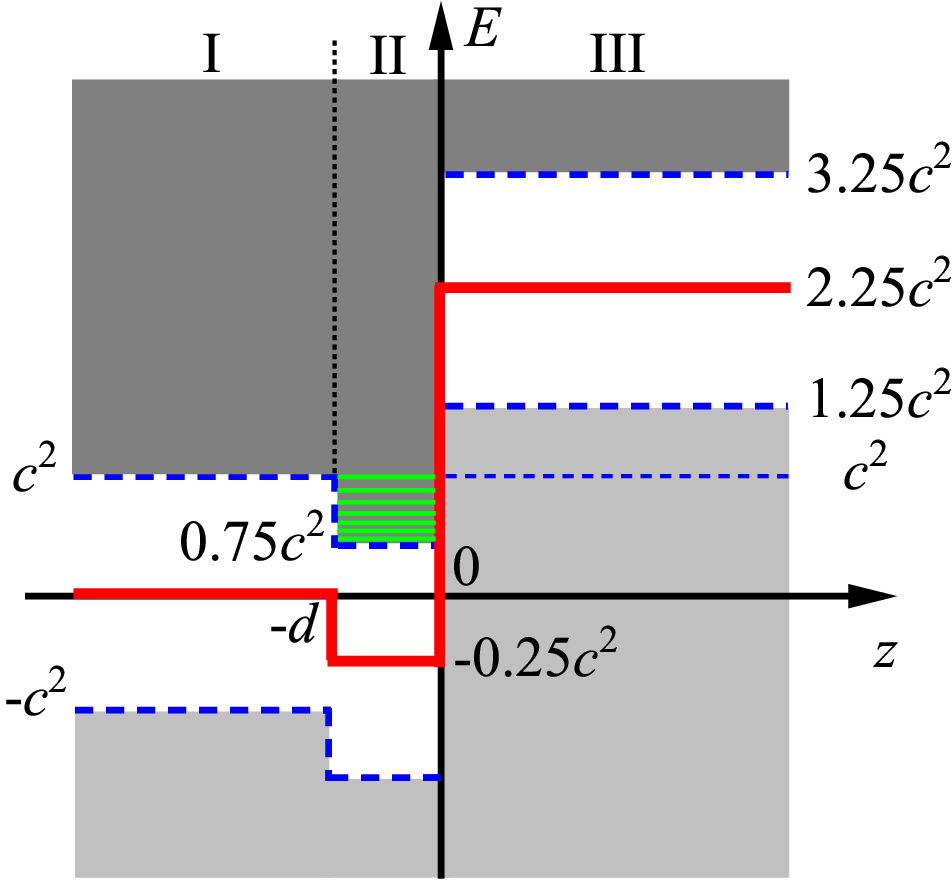}%
\caption{Sketch of energy levels for an asymmetric potential well (thick red line). The resonant state energy levels (green bar) are shown in the potential well. I, II, and III represent there regions divided by the potential. The parameters of a typical potential are $V_1=2.5c^2$ and $V_2=0.25c^2$.}
\label{fig:asy-potential-well}
\end{figure}

The one-dimensional asymmetric potential well (\ref{eqn:Potential}) with $w\rightarrow0$ divides the space into three regions as
\begin{equation}
-V(z)=\left\{\begin{array}{cccc}0\;,&z<-d&\text{(region}&\text{I)}\\
-V_2\;,&-d\leq z\leq0&\text{(region}&\text{II)}\\
V_1-V_2\;,&z>0&\text{(region}&\text{III)}\end{array}\right.
\end{equation}
see Fig. \ref{fig:asy-potential-well}.
The Dirac equation in these three regions reads
\begin{equation}
\begin{aligned}
&\text{I:}\quad(\hat{\alpha}\cdot\hat{p}c+\hat{\beta}c^{2})\psi=E\psi, \quad z<-d\\
&\text{II:}\quad(\hat{\alpha}\cdot\hat{p}c+\hat{\beta}c^{2})\psi
=\begin{pmatrix}E+V_{2}\end{pmatrix}\psi, \quad-d\leq z\leq0\\
&\text{III:}\quad(\hat{\alpha}\cdot\hat{p}c+\hat{\beta}c^{2})\psi
=\begin{pmatrix}E+V_{2}-V_{1}\end{pmatrix}\psi, \quad z>0\end{aligned}
\end{equation}
The solutions of the above Dirac equation with spin up are given as
\begin{equation}
\begin{split}
\psi_\mathrm{I}(z)=Ae^{\mathrm{i}p_1z}\begin{pmatrix}1\\0\\
\frac{p_1c}{E+c^2}\\0\end{pmatrix}
+A'e^{-\mathrm{i}p_1z}\begin{pmatrix}1\\0\\ \frac{-p_1c}{E+c^2}\\0
\end{pmatrix},
\end{split}
\end{equation}
where $p_1^2=\frac{E^2}{c^2}-c^2$;
\begin{equation}
\begin{split}
\psi_\mathrm{II}(z)=Be^{\mathrm{i}p_2z}\begin{pmatrix}1\\0\\
\frac{p_2c}{E+V_2+c^2}\\0\end{pmatrix}
+B'e^{-\mathrm{i}p_2z}\begin{pmatrix}1\\0\\ \frac{-p_2c}{E+V_2+c^2}\\0
\end{pmatrix}\!,
\end{split}
\end{equation}
where $p_2^2=\frac{(E+V_2)^2}{c^2}-c^2$;
\begin{equation}
\begin{split}
&\psi_\mathrm{III}(z)=\\&Ce^{\mathrm{i}p_3z}\begin{pmatrix}1\\0\\
\frac{p_3c}{E+V_2-V_1+c^2}\\0\end{pmatrix}
+C'e^{-\mathrm{i}p_3z}\begin{pmatrix}1\\0\\ \frac{-p_3c}{E+V_2-V_1+c^2}\\0
\end{pmatrix},
\end{split}
\end{equation}
where $p_3^2=\frac{(E+V_2-V_1)^2}{c^2}-c^2$.

According to the continuity conditions of the wave function at the boundaries of the potential well, $\psi_\mathrm{I}(-d)=\psi_\mathrm{II}(-d)$ and $\psi_\mathrm{II}(0)=\psi_\mathrm{III}(0)$,
we get
\begin{equation}\label{eqn:AB}
\begin{pmatrix}A\\A'\end{pmatrix}=\frac{1}{2}\Bigg(\begin{array}{ccc}
\frac{\gamma+1}{\gamma}e^{\mathrm{i}(p_1-p_2)d}&\frac{\gamma-1}{\gamma}
e^{\mathrm{i}(p_1+p_2)d}\\\frac{\gamma-1}{\gamma}e^{-\mathrm{i}(p_1+p_2)d}
&\frac{\gamma+1}{\gamma}e^{\mathrm{i}(p_2-p_1)d}\end{array}\Bigg)
\Bigg(\begin{array}{c}B\\B'\end{array}\Bigg)
\end{equation}
and
\begin{equation}\label{eqn:BC}
\binom{B}{B'}=\frac{1}{2}\Bigg(\begin{matrix}\frac{\tau+1}{\tau}
&\frac{\tau-1}{\tau}\\\frac{\tau-1}{\tau}&\frac{\tau+1}{\tau}
\end{matrix}\Bigg)\Bigg(\begin{matrix}C\\C'\end{matrix}\Bigg),
\end{equation}
where
\begin{equation}\label{eqn:gammatau}
\begin{split}
\gamma=&\frac{p_1c}{E+c^2}\frac{E+V_2+c^2}{p_2c},\\
\tau=&\frac{p_2c}{E+V_2+c^2}\frac{E+V_2-V_1+c^2}{p_3c}.
\end{split}
\end{equation}
From Eqs. (\ref{eqn:AB}) and (\ref{eqn:BC}), we have
\begin{widetext}
\begin{equation}\label{eqn:AC}
\left(\begin{array}{c}A\\A'\end{array}\right)
=\frac{1}{4}\left(\begin{array}{cc}\frac{\gamma+1}{\gamma}\frac{\tau+1}{\tau}
e^{\mathrm{i}(p_1-p_2)d}+\frac{\gamma-1}{\gamma}\frac{\tau-1}{\tau}
e^{\mathrm{i}(p_1+p_2)d}&\frac{\gamma+1}{\gamma}\frac{\tau-1}{\tau}
e^{\mathrm{i}(p_1-p_2)d}+\frac{\gamma-1}{\gamma}\frac{\tau+1}{\tau}
e^{\mathrm{i}(p_1+p_2)d}\\
\frac{\gamma-1}{\gamma}\frac{\tau+1}{\tau}
e^{-\mathrm{i}(p_1+p_2)d}+\frac{\gamma+1}{\gamma}\frac{\tau-1}{\tau}
e^{\mathrm{i}(p_2-p_1)d}&\frac{\gamma-1}{\gamma}\frac{\tau-1}{\tau}
e^{-\mathrm{i}(p_1+p_2)d}+\frac{\gamma+1}{\gamma}\frac{\tau+1}{\tau}
e^{\mathrm{i}(p_2-p_1)d}\end{array}\right)
\left(\begin{array}{c}C\\C'\end{array}\right).
\end{equation}
\end{widetext}
For the bound states in the asymmetric potential well, $A$ and $C'$ must be zero, so the first of (\ref{eqn:AC}) becomes
\begin{equation}
0=\frac{C}{4}\Big[\frac{\gamma+1}{\gamma}\frac{\tau+1}{\tau}
e^{\mathrm{i}(p_1-p_2)d}+\frac{\gamma-1}{\gamma}\frac{\tau-1}{\tau}
e^{\mathrm{i}(p_1+p_2)d}\Big].
\end{equation}
Since $C\neq0$, we obtain
\begin{equation}
(\gamma+1)(\tau+1)e^{-\mathrm{i}p_2d}+(\gamma-1)(\tau-1)e^{\mathrm{i}p_2d}=0
\end{equation}
or
\begin{equation}
\begin{split}
&(\gamma+1)(\tau+1)[\cos(p_2d)-\mathrm{i}\sin(p_2d)]\\
&+(\gamma-1)(\tau-1)[\cos(p_2d)+\mathrm{i}\sin(p_2d)]=0.
\end{split}
\end{equation}
Finally, we get the formula for calculating bound state energy levels in an asymmetric potential well:
\begin{equation}\label{eqn:bsel}
\mathrm{i}\tan(p_2d)=\frac{\gamma\tau+1}{\gamma+\tau}.
\end{equation}
For a symmetric potential well, the potential height $V_1-V_2$ in region III is zero. So we can obtain $\tau=1/\gamma$ from Eq. (\ref{eqn:gammatau}). Then equation (\ref{eqn:bsel}) becomes
\begin{equation}\label{eqn:bsel1}
\mathrm{i}\tan(p_2d)=\frac{2\gamma}{1+\gamma^2}.
\end{equation}
This is just the formula for calculating bound state energy levels in a symmetric potential well.

\begin{figure}[!ht]
\centering
\includegraphics[width=0.45\textwidth]{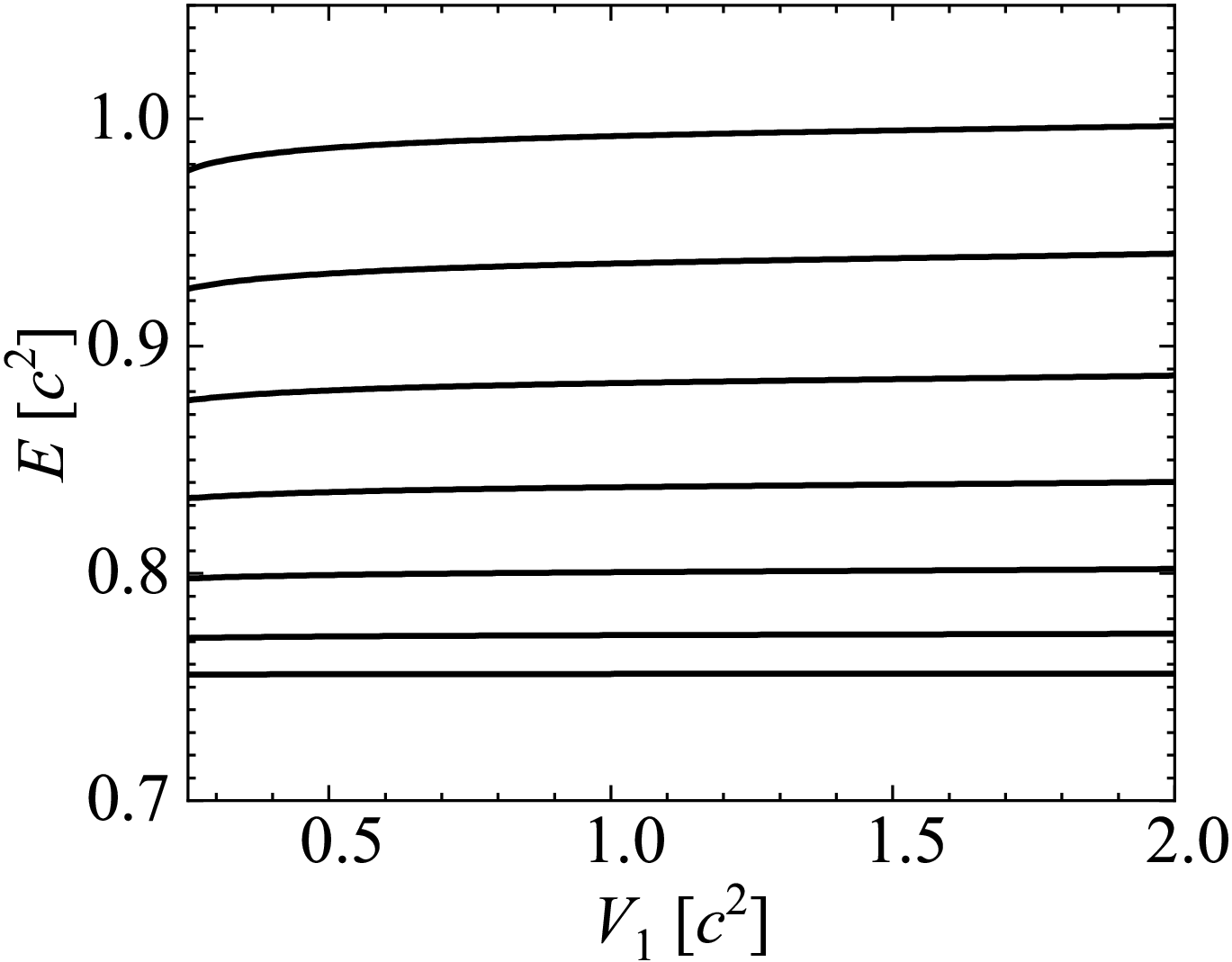}%
\caption{The variation of bound state energy with the potential height $V_1$ calculated according to Eq. (\ref{eqn:bsel}). Other potential parameters are $V_2=0.25c^2$, $w\rightarrow0$, and $d=\,0.2\mathrm{a.u.}$.}
\label{fig:E-V1}
\end{figure}

In Fig. \ref{fig:E-V1}, we plot the variation of bound state energy with the potential height $V_1$ according to Eq. (\ref{eqn:bsel}). One can see that there are seven bound states in the asymmetric potential well and their energy grows with the increase of $V_1$. When the potential height $V_1$ is large, the change of bound state energy with the potential height is relatively gentle, and there is an approximate linear relationship between them.
We fit the relationship between them based on the data within the range where $V_1$ is greater than $c^2$ and less than $2c^2$.
The fitting functions and correlation coefficients are shown in Table \ref{tab:E-V1}.
From the results, we do see a strong linear relationship between the bound state energy and the potential height for large values of the potential height.
Assuming that the fitting functions still hold for $V_1>2c^2$, we can compute the bound state energy in the supercritical asymmetric potential well with $V_1=2.5c^2$.
The results are shown in the second column of Table \ref{tab:comparison1} and compared with the results of CQFT.
It can be observed that the maximum relative error of the results obtained by the two methods is less than 0.1\%.
On one hand, this demonstrates the feasibility of our fitting method; on the other hand, it confirms that the discrete peaks in the positron energy spectrum indeed arise from the overlap between bound states and the negative energy continuum.
Notice that our fitting method is feasible because when the bound states dive into the negative energy continuum, their energy is dominantly linear in $V_1-2c^2$ with some smaller deviations quadratic in $V_1-2c^2$ \cite{Muller1972}. In CQFT, the energy of resonant states (i.e., the bound states overlapping with negative energy continuum) is calculated by $V_1-V_2-E_{\mathrm{peak}}$, where $E_{\mathrm{peak}}$ is the energy corresponding to the discrete peaks in the positron energy spectrum.

\begin{table}[!ht]\suppressfloats
\renewcommand{\arraystretch}{1.5}
\setlength{\tabcolsep}{5pt}
\begin{center}
   \caption{\label{tab:E-V1} The fitting functions and correlation coefficients between bound state energy and the larger potential height $V_1$. Other potential parameters are $V_2=0.25c^2$, $w\rightarrow0$, and $d=\,0.2\mathrm{a.u.}$.}
   \begin{tabular}{ccc} \hline \hline
    \makecell[c]{Bound state \\number}  & Fitting functions&   \makecell[c]{Correlation \\coefficients}   \\ \hline
    $1$ & $E=0.98831+0.00435V_1$ & $0.99792$  \\ 
    $2$ & $E=0.93232+0.00423V_1$ & $0.99913$  \\ 
    $3$ & $E=0.88057+0.00326V_1$ & $0.99944$ \\
    $4$ & $E=0.83568+0.00227V_1$ & $0.99946$  \\
    $5$ & $E=0.79916+0.00138V_1$ & $0.99913$  \\
    $6$ & $E=0.77216+0.000651V_1$ & $0.99834$ \\ 
    $7$ & $E=0.75559+0.000170V_1$ & $0.99711$  \\ 
     \hline \hline
   \end{tabular}
   \end{center}
\end{table}

In Appendix \ref{appa}, we illustrate that the bound state energy level formula (\ref{eqn:bsel1}) for a subcritical symmetric potential well is also applicable to calculate resonant state energy in a supercritical symmetric potential well.
Since the solutions of this formula for the supercritical symmetric potential well are complex, it can not only determine the energy of the resonant states (corresponding to the real part of the solutions), but also provide the width of the resonant states (corresponding to the imaginary part of the solutions).
Similarly, we can directly apply the bound state energy level formula (\ref{eqn:bsel}) for the subcritical asymmetric potential well to the complex energy region to predict the energy and width of resonant states in the supercritical asymmetric potential well.
By comparing the results from the analytical formula with those from CQFT, see the last column of Table \ref{tab:comparison1}, we find that the maximum relative error is less than $0.13\%$.
The formula (\ref{eqn:bsel}) can effectively estimate the resonant state energy in the supercritical asymmetric potential well.
This result not only shows the feasibility of extending the bound state energy level formula (\ref{eqn:bsel}) into the complex energy region, but also validates that the discrete peaks in the energy spectrum of positrons created in a supercritical asymmetric potential well do result from the formation of resonant states.
The reason why the formula (\ref{eqn:bsel}) can be extended to the complex energy region to calculate the resonant state energy is that a resonant state itself is a degenerate state of a bound state and the negative energy continuum, and inherently retains some bound state characteristics.
Furthermore, the imaginary part of the resonant state energy computed by formula (\ref{eqn:bsel}) is related to the pair creation rate \cite{Lv2013,Lv2014,Zhou2023}.

\begin{table*}[!ht]\suppressfloats
\renewcommand{\arraystretch}{1.5}
\setlength{\tabcolsep}{9pt}
\begin{center}
   \caption{\label{tab:comparison1} Comparison of the resonant state energy computed by fitting functions ($E_\mathrm{fitt}$), the bound state energy level formula ($E_\mathrm{boud}$), and the CQFT at $t=0.03\,\mathrm{a.u.}$ ($E_\mathrm{CQFT}$). The potential width $w$ approaches to $0$ for the first two methods and equals $0.075/c$ for CQFT. Other potential parameters are $V_1=2.5c^2,\,V_2=0.25c^2$, and $d=\,0.2\mathrm{a.u.}$.}
   \begin{tabular}{cccccc} \hline \hline
    \makecell[c]{Resonant state \\number} & $E_\mathrm{fitt}\,\,[\mathrm{c^2}]$ & $ E_\mathrm{boud}\,\,[\mathrm{c^2}]$ & $ E_\mathrm{CQFT}\,\,[\mathrm{c^2}]$ & $\big|\frac{E_\mathrm{fitt}-E_\mathrm{CQFT}}{E_\mathrm{CQFT}}
    \big|\,\,[\%]$&  $\big|\frac{\Re(E_\mathrm{boud})-E_\mathrm{CQFT}}{E_\mathrm{CQFT}}
    \big|\,\,[\%]$ \\ \hline
    $1$ & 0.999185 & $0.999983+0\mathrm{i}$ & $0.999517$ & 0.0332 & $0.0467$ \\ 
    $2$ & 0.942895 & $0.942485+0.001966\mathrm{i}$ & $0.942401$ & 0.0524 & $0.0089$ \\ 
    $3$ & 0.888720 & $0.888226+0.001625\mathrm{i}$ & $0.889318$ & 0.0672 & $0.1228$ \\
    $4$ & 0.841355 & $0.840904+0.001185\mathrm{i}$ & $0.841770$ & 0.0493 & $0.1029$ \\
    $5$ & 0.802610 & $0.802255+0.000737\mathrm{i}$ & $0.802918$ & 0.0384 & $0.0826$ \\
    $6$ & 0.773788 & $0.773598+0.000351\mathrm{i}$ & $0.773659$ & 0.0166 & $0.0079$ \\ 
    $7$ & 0.756015 & $0.755958+0.000092\mathrm{i}$ & $0.756713$ & 0.0922 & $0.0998$ \\ 
     \hline \hline
   \end{tabular}
   \end{center}
\end{table*}


\subsection{Effect of the subcritical potential height on the energy spectra of created particles} \label{sec:sub3b}

We fix the potential well height $V_1 = 2.5c^2$ and vary $V_2$ from $0$ to $0.5c^2$ to study the effect of the subcritical potential height on pair creation.
For $V_1 = 2.5c^2$ and $V_2 = 0$, the only overlap occurs between the positive and negative energy continuum.
When $V_2$ is non-zero, bound states begin to overlap with the negative energy continuum.
A further increase in $V_2$ widens the energy range over which bound states overlap with the negative energy continuum, whereas the overlap region between the positive and negative energy continuum shrinks.
At $V_2=0.5c^2$, the system is completely governed by the bound -continuum overlap, with no remaining continuum-continuum overlap.
Figure \ref{fig:asy-potential-well} displays the sketch of energy levels for the asymmetric potential well, which can help us understand the impact of $V_2$ on energy levels. For $V_1 = 2.5c^2$ and $V_2 = 0.25c^2$, there are seven resonant states in the supercritical asymmetric potential well.

\begin{figure}[!ht]
\centering
\begin{minipage}{0.9\linewidth}
\centering
\includegraphics[width=\textwidth]{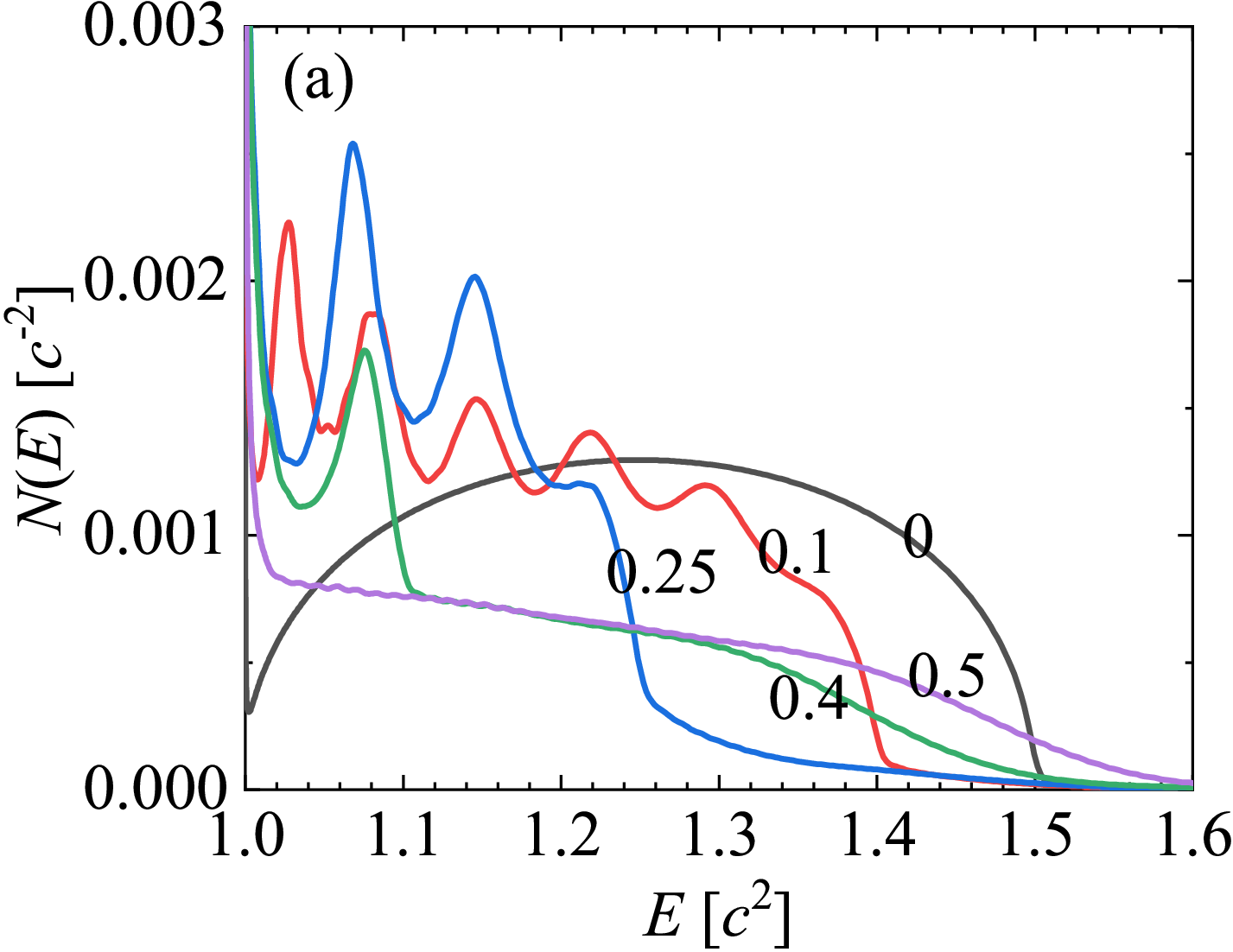}%
\end{minipage}
\begin{minipage}{0.9\linewidth}
\centering
\includegraphics[width=\textwidth]{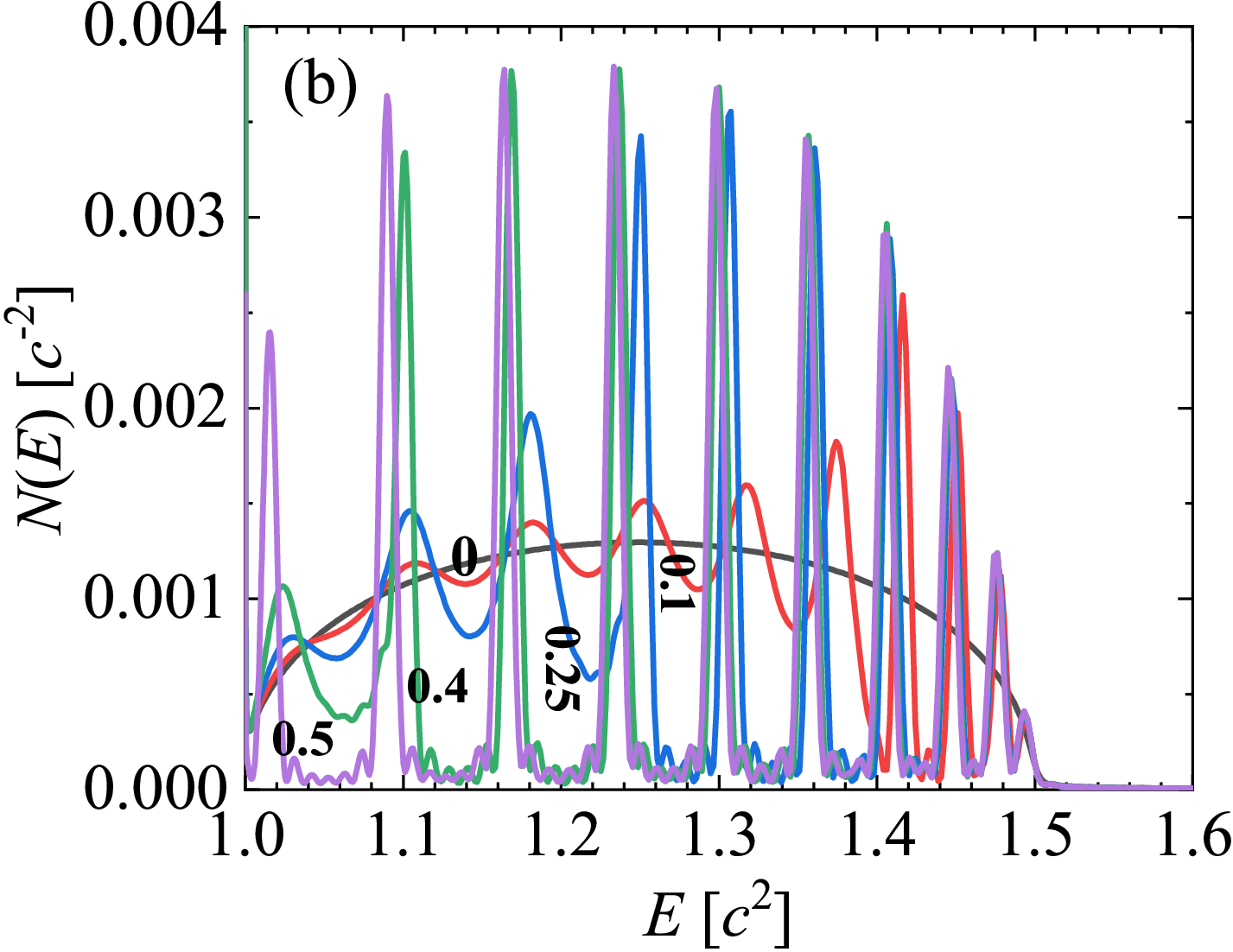}%
\end{minipage}
\centering
\caption{Energy spectra of created electrons (a) and positrons (b) at $t=0.03\,\mathrm{a.u.}$ for $V_2=0,\,0.1,\,0.25,\,0.4,\,0.5c^2$. Each line is marked with the value of $V_2$. Other potential parameters are $V_1=2.5c^2, w=0.3/c, d=0.2\,\mathrm{a.u.}, N_z=8192, L=16\,\mathrm{a.u.}$.}
\label{fig:varying-v2-NE-pe}
\end{figure}

Using CQFT, we compute the energy spectra of electrons and positrons created in supercritical asymmetric potential wells, as shown in Fig. \ref{fig:varying-v2-NE-pe}.
The results reveal a clear breaking of symmetry between the electron and positron energy spectra, along with discrete peaks arising in the high-energy part of the positron spectrum.
These features are primarily related to the presence of the potential well.
The low-energy portion of the positron spectrum corresponds to electron-positron pair creation from the continuum-continuum overlap, while the discrete peaks are linked to pair creation that results from the bound-continuum overlap.

In addition, from the electron energy spectrum shown in Fig. \ref{fig:varying-v2-NE-pe}(a), it can observed that with the increase of $V_2$ from $0$ to $0.5c^2$, oscillatory structures gradually emerge in the spectrum.
These oscillations are induced by scattering resonances above the potential well.
Moreover, as $V_2$ increases, the electron energy spectrum first becomes more concentrated, reaching maximum concentration at $V_2 = 0.25c^2$, and then becomes more dispersed.
This indicates that there exists an optimal subcritical potential height that maximizes the energy concentration of electrons created in a supercritical asymmetric potential well.
This result can be understood as follows: when the subcritical potential height is $0.25c^2$, the energy range of the bound-continuum overlap is exactly equal to that of the continuum-continuum overlap; this energy range is half of that of the total overlap region.
At this point, the energy of electrons created in these two overlap regions falls within the range of $c^2$ to $1.25c^2$.
The energy spectrum is most concentrated.
If the energy range of these two overlap regions is not equal, there must be an overlap region that produces electrons with energy beyond the range of $c^2$ to $1.25c^2$.
Therefore, the electron energy spectrum is more dispersed.
For instance, when $V_2 = 0.4c^2$, the energy range of electrons created in continuum-continuum overlap is $c^2$ to $1.1c^2$, while the energy range of electrons created in the bound-continuum overlap is $c^2$ to $1.4c^2$.
Hence, the total energy range of created electrons is $c^2$ to $1.4c^2$, which is more larger than that in the case of $V_2 = 0.25c^2$.
This indicates that the energy spread of electrons created in a supercritical potential can be controlled by adjusting the subcritical potential height.
For the positron energy spectrum shown in Fig. \ref{fig:varying-v2-NE-pe}(b), different from the creation of electrons, the creation of positrons caused by the bound-continuum overlap (discrete peaks) is clearly separated from that induced by the continuum-continuum overlap (broad peaks).
This is because positrons created by the bound-continuum overlap are accelerated away from the potential well instead of being confined in the well like electrons.
With the increase of $V_2$, the overlap region between bound states and the negative energy continuum becomes large and more peaks appear.
When $V_2=0.5c^2$, there are no broad peaks in the spectrum. All positrons are produced by the bound-continuum overlap.

\subsection{Time evolution of created particles in the asymmetric potential well}\label{sec:sub3c}

In this subsection, we will numerically compute the time evolution of particle number and the energy spectrum of created particles to study pair creation in supercritical step potentials (only having the continuum-continuum overlap) and asymmetric potential wells (including both the bound-continuum overlap and the continuum-continuum overlap).
This allows us to explore the mechanisms of pair creation when both bound states and the positive energy continuum overlap with the negative energy continuum.
To better understand the time evolution of particles created in bound-continuum and continuum-continuum overlap regions, we study the number and energy spectrum of created positrons rather than electrons in this subsection.

\begin{figure}[!ht]
\centering
\includegraphics[width=0.45\textwidth]{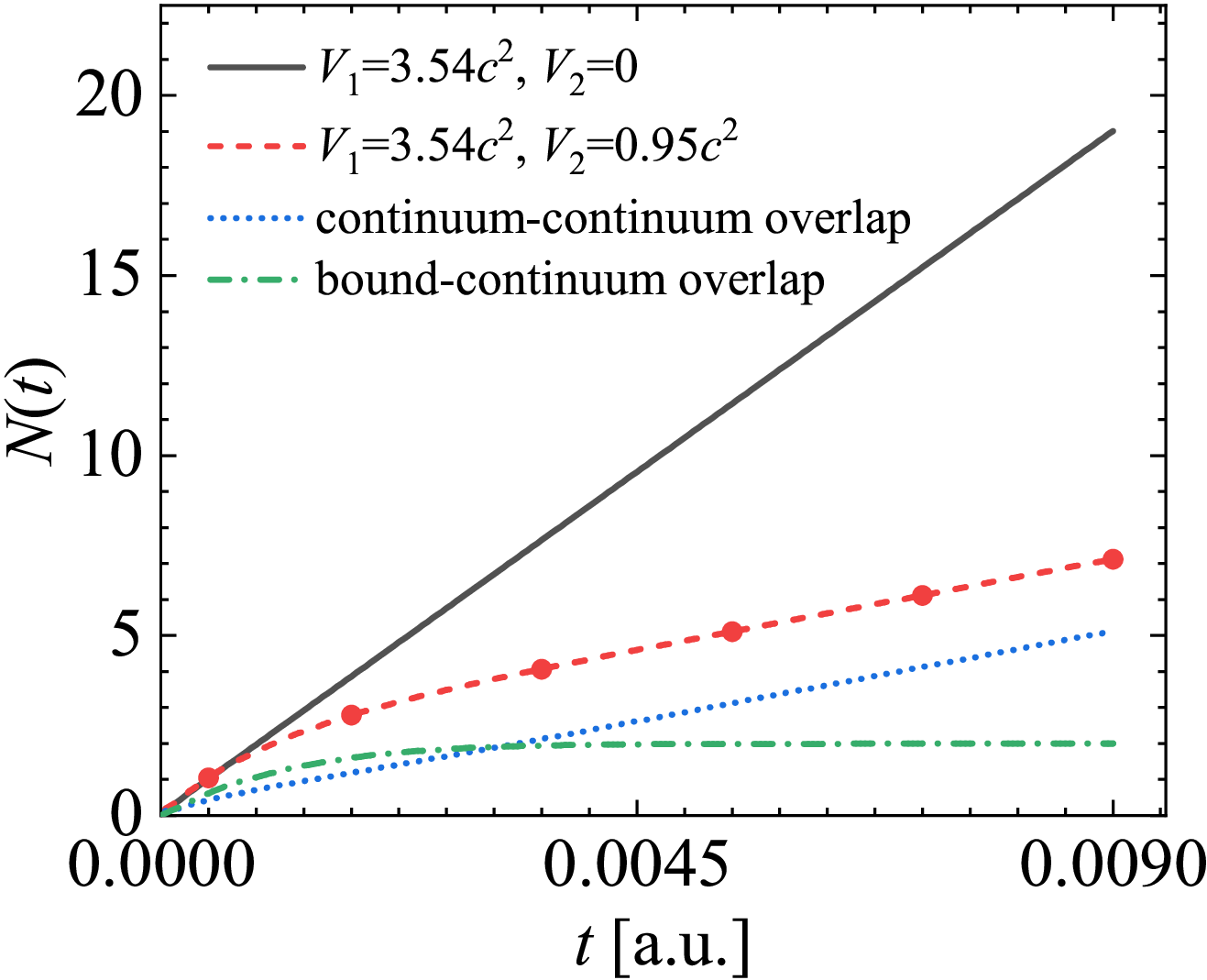}%
\caption{Time evolution of the number of positrons created in the step potential with $V_1=3.54c^2, V_2=0$ (black solid line), the asymmetric potential well with $V_1=3.54c^2, V_2=0.95c^2$ (red dashed line), the overlap region between the positive and negative continuum in the asymmetric potential well (blue dotted line), and the overlap region between bound sates and the negative continuum in the asymmetric potential well (green dash-dotted line). Other parameters are $w=0.3/c, d=0.2\,\mathrm{a.u.}, N_z=2048, L=8\, \mathrm{a.u.}$.}
\label{fig:Evolution-Nt}
\end{figure}

\begin{figure}[!ht]
\centering
\begin{minipage}{0.49\linewidth}
\centering
\includegraphics[width=\textwidth]{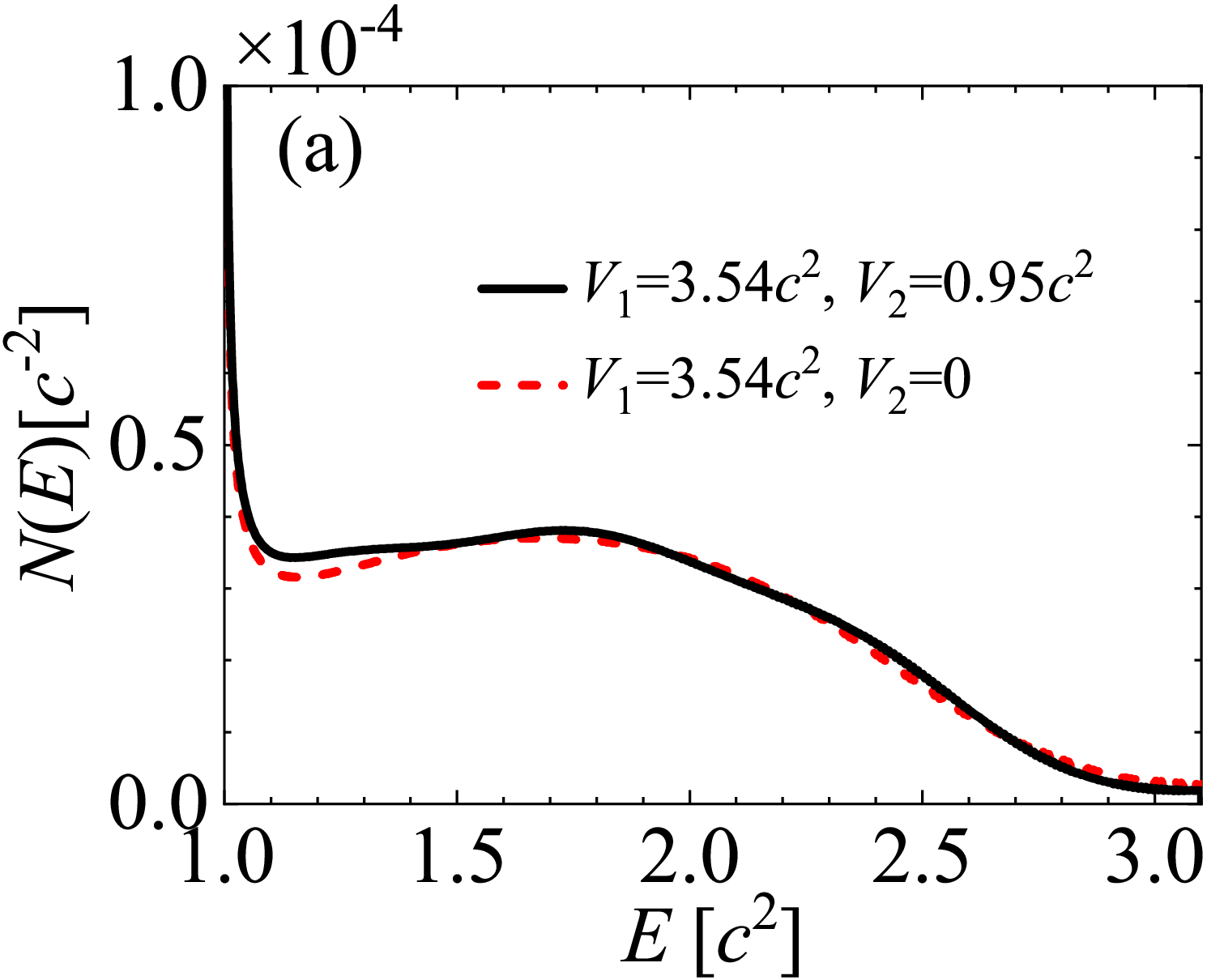}%
\end{minipage}
\begin{minipage}{0.49\linewidth}
\centering
\includegraphics[width=0.938\textwidth]{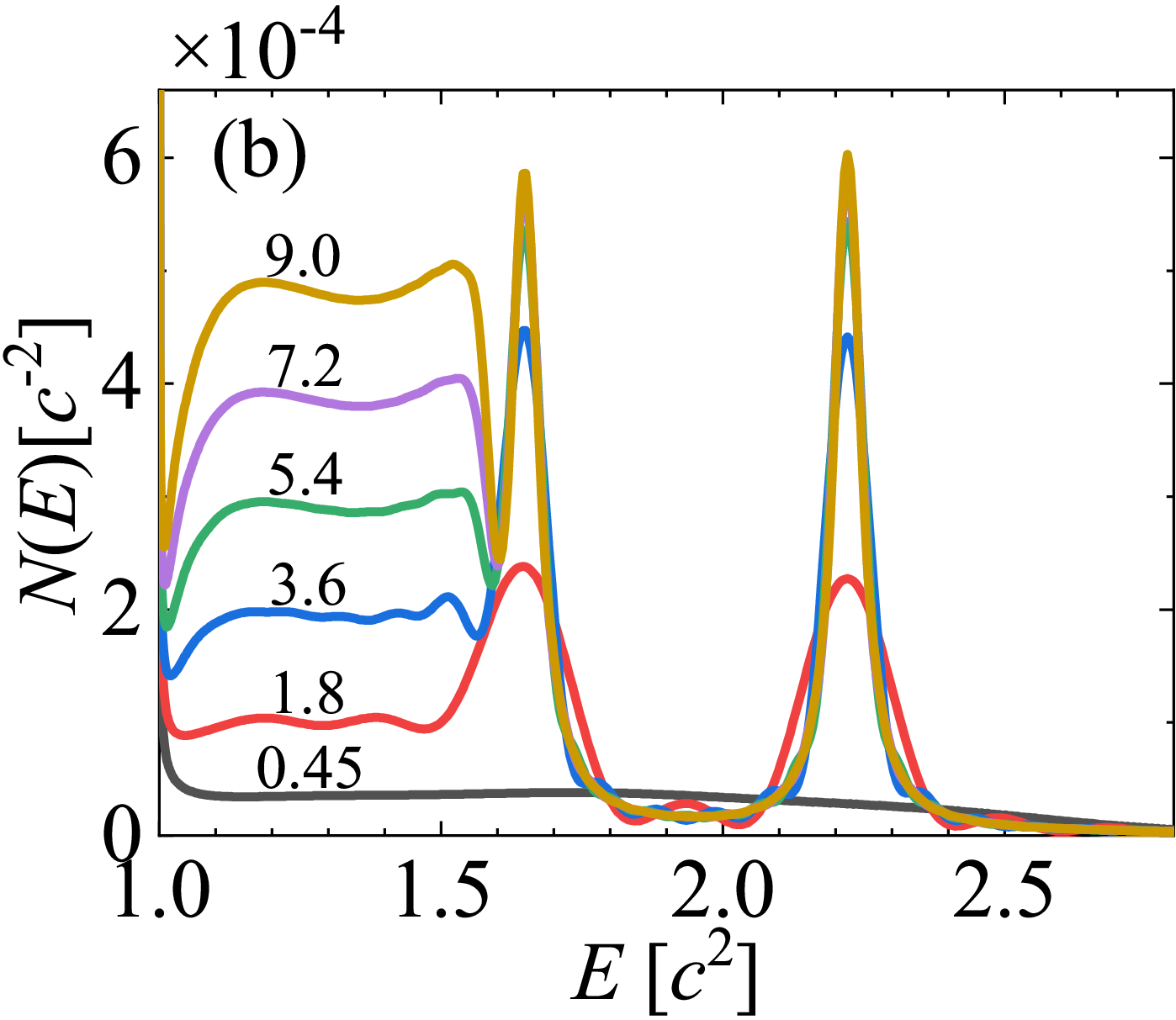}%
\end{minipage}
\centering
\caption{Energy spectra of created positrons at different times. (a) shows the comparison of energy spectra for the asymmetric potential well ($V_1=3.54c^2, V_2=0.95c^2$) and the step potential ($V_1=3.54c^2, V_2=0$) at $t=4.5\times10^{-4}\,\mathrm{a.u.}$. (b) shows the energy spectra for the asymmetric potential well at $t=0.45,\,1.8,\,3.6,\,5.4,\,7.2,\,9.0\times10^{-3}\,\mathrm{a.u.}$ (corresponding to red points in Fig.\ref{fig:Evolution-Nt}). The time values are marked on the lines. Other parameters are the same as in Fig.\ref{fig:Evolution-Nt}.}
\label{fig:Evolution-NE0t}
\end{figure}

The height of the asymmetric potential well we studied is $V_1 = 3.54c^2$ and $V_2 = 0.95c^2$.
The profile of the potential well is similar to Fig. \ref{fig:asy-potential-well}.
In the energy range of $0.05c^2$ to $c^2$, the bound states in the well overlap with the negative energy continuum, while in the energy range of $c^2$ to $1.59c^2$, the positive energy continuum overlaps with the negative energy continuum.
Figure \ref{fig:Evolution-Nt} displays the time evolution of the particle number.
The red dashed line represents the time evolution of the number of positrons produced in the asymmetric potential well with $V_1 = 3.54c^2$ and $V_2 = 0.95c^2$.
In this potential well, there are two bound states that dive into the negative energy continuum and form two resonant states.
The black solid line shows the time evolution of the number of positrons created in a step potential with $V_1= 3.54c^2$ and no subcritical potential (i.e., $V_2=0$).
The blue dotted line is the time evolution of the number of positrons created by the continuum-continuum overlap in the asymmetric potential well with $V_1 = 3.54c^2$ and $V_2 = 0.95c^2$.
The green dash-dotted line is the time evolution of the number of positrons created by the bound-continuum overlap.
From Fig. \ref{fig:Evolution-Nt}, one can see that in the initial stage of positron creation in the asymmetric potential well, the growth trend of the positron number over time is the same as that in the step potential.
This is because in the very early initial period, electrons produced by the supercritical potential have not yet reached the potential well boundary where the subcritical potential is located, and thus are not affected by the subcritical potential.
Consequently, the pair creation in the asymmetric potential well is similar to that in the supercritical step potential.
This is verified in Fig. \ref{fig:Evolution-NE0t}(a), where the initial positron energy spectrum for the asymmetric potential well is compared with that for the step potential.
Notice that the energy spectrum of electrons is the same as that of positrons in this initial period.
Subsequently, the number of positrons created in the asymmetric potential well grows as a cubic polynomial over time.
In the long-time limit, it grows linearly with a constant rate.
The reason is that once the bound state energy levels are filled with electrons, reaching the number of bound states, the Pauli exclusion principle prevents further pair creation by the bound-continuum overlap, see the green dash-dotted line.
At this time, only the overlap region between the positive and negative energy continuum continues to produce positrons at a constant rate.
This result is obtained by comparing the growth rate of red dashed line and the blue dotted line.

In Fig. \ref{fig:Evolution-NE0t}, the positron energy spectra at different times are depicted.
At the initial evolution time $t_1 = 4.5 \times 10^{-4}$ a.u., the positron energy spectrum for the asymmetric potential well with $V_1 = 3.54c^2, V_2 = 0.95c^2$ (black solid line) is smooth and dispersed.
It nearly identical to the energy spectrum of positrons created in the step potential with $V_1 = 3.54c^2, V_2 = 0$, see red dashed line in \ref{fig:Evolution-NE0t}(a).
As time grows, the number of created positrons increases greatly in both overlap regions, see \ref{fig:Evolution-NE0t}(b).
The positron energy spectrum corresponding to the continuum-continuum overlap remains dispersed, while the positron energy spectrum corresponding to the bound-continuum overlap exhibits discrete peaks at $E = 1.65c^2$ and $2.22c^2$.
In the long-time limit, due to the Pauli exclusion principle, the growth of the positron number corresponding to the two peaks gradually slows down, and eventually stops.
However, the overlap region between the positive and negative energy continuum continues to produce positrons at a high rate. This result further confirms our earlier analysis of the time evolution of pair creation in the supercritical asymmetric potential well.

By comparing the growth rate of particles created in a supercritical asymmetric potential well (including the bound-continuum and continuum-continuum overlap) with those created in the potentials where only one type of overlap exists, we explore a new way for enhancing pair production.

\begin{figure}[!ht]
\centering
\includegraphics[width=0.48\textwidth]{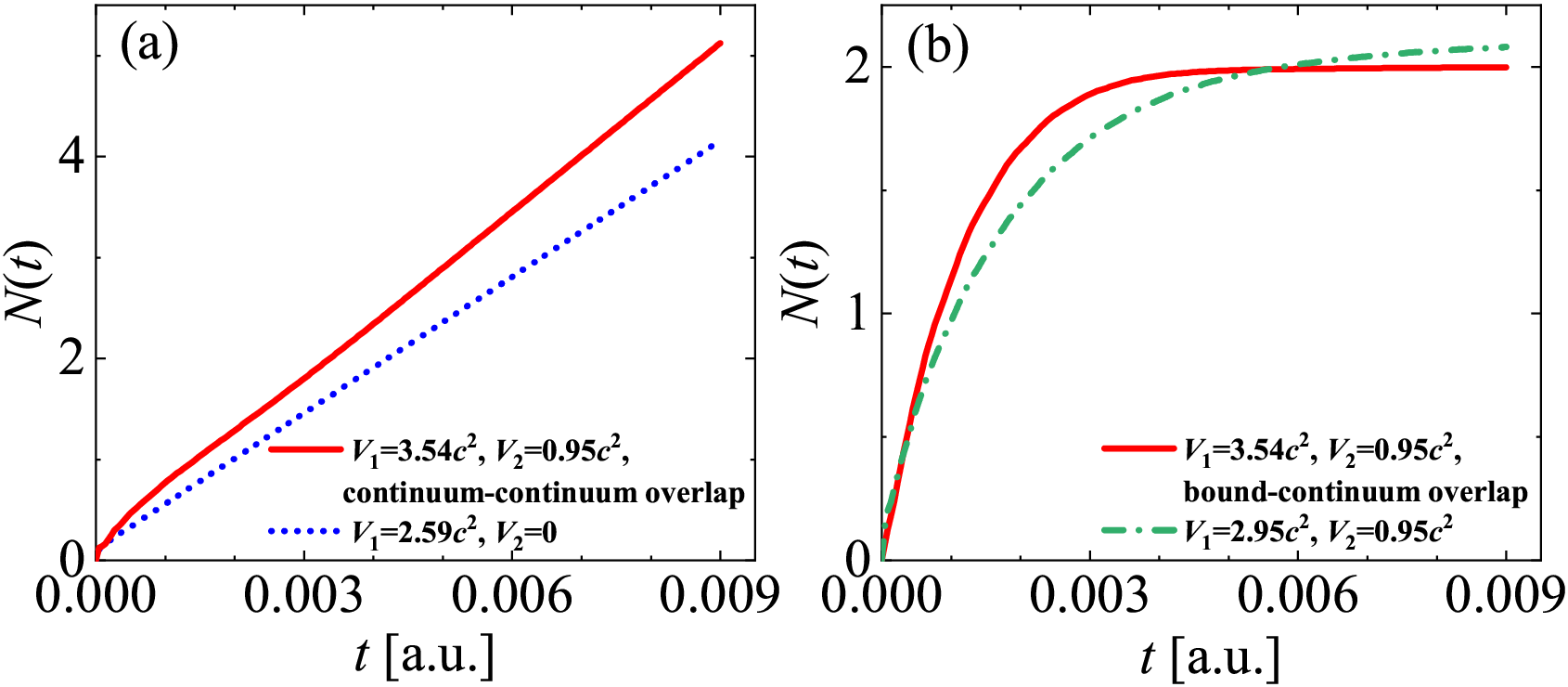}%
\caption{Time evolution of the number of positrons created by the continuum-continuum overlap (red solid line in (a)) and bound-continuum overlap (red solid line in (b)) in the asymmetric potential well with $V_1=3.54c^2, V_2=0.95c^2$. The results are compared with those for a step potential with $V_1=2.59c^2, V_2=0$ (blue dotted line in (a)) and an asymmetric potential well with $V_1=2.95c^2, V_2=0.95c^2$ (green dash-dotted line in (b)). Other parameters are $w=0.3/c, d=0.2\,\mathrm{a.u.}, N_z=2048, L=8\,\mathrm{a.u.}$.}
\label{fig:Ehancement-Nt-cb}
\end{figure}

In Fig. \ref{fig:Ehancement-Nt-cb}, we plot the time evolution of the number of positrons created by the continuum-continuum overlap and bound-continuum overlap in the asymmetric potential well with $V_1 = 3.54c^2, V_2 = 0.95c^2$, and compare them with the cases where only the continuum-continuum overlap or bound-continuum overlap exists.
In Fig. \ref{fig:Ehancement-Nt-cb}(a), the red solid line corresponds to the positrons produced by the continuum-continuum overlap in the asymmetric potential well with $V_1 = 3.54c^2, V_2 = 0.95c^2$.
The blue dotted line represents the positrons created in a step potential with $V_1= 2.59c^2, V_2=0$, where only the continuum-continuum overlap exists and the energy range of the overlap region matches that in the supercritical asymmetric potential well.
We find that the growth rate of the number of positrons created in by the continuum-continuum overlap is higher in the asymmetric potential well than in the step potential.
In Fig. \ref{fig:Ehancement-Nt-cb}(b), the red solid line shows the positrons produced by the bound-continuum overlap in the asymmetric potential well with $V_1 = 3.54c^2, V_2 = 0.95c^2$.
The green dash-dotted line corresponds to the positrons produced in an asymmetric potential well with $V_1 = 2.95c^2, V_2 = 0.95c^2$, where only the bound-continuum overlap exists and the energy range of this overlap region matches that in the supercritical asymmetric potential well.
We observe that, in the long-time limit, the positron number approaches a constant value due to the Pauli exclusion principle.
Moreover, the constant value for the asymmetric potential well with $V_1 = 2.95c^2, V_2 = 0.95c^2$ is larger than that for the asymmetric potential well with $V_1 = 3.54c^2, V_2 = 0.95c^2$.
The above results indicate that electron-positron pair creation in a supercritical asymmetric potential well (having the continuum-continuum and bound-continuum overlap) is not a simple combination of pair creation induced by two types of overlap separately.
In fact, this can also be seen from the oscillatory structures of the energy spectra in Fig. \ref{fig:varying-v2-NE-pe}(a), because there is no oscillation in the energy spectrum when only the continuum-continuum overlap exists, see the black solid line.

\begin{figure}[!ht]
\centering
\includegraphics[width=0.45\textwidth]{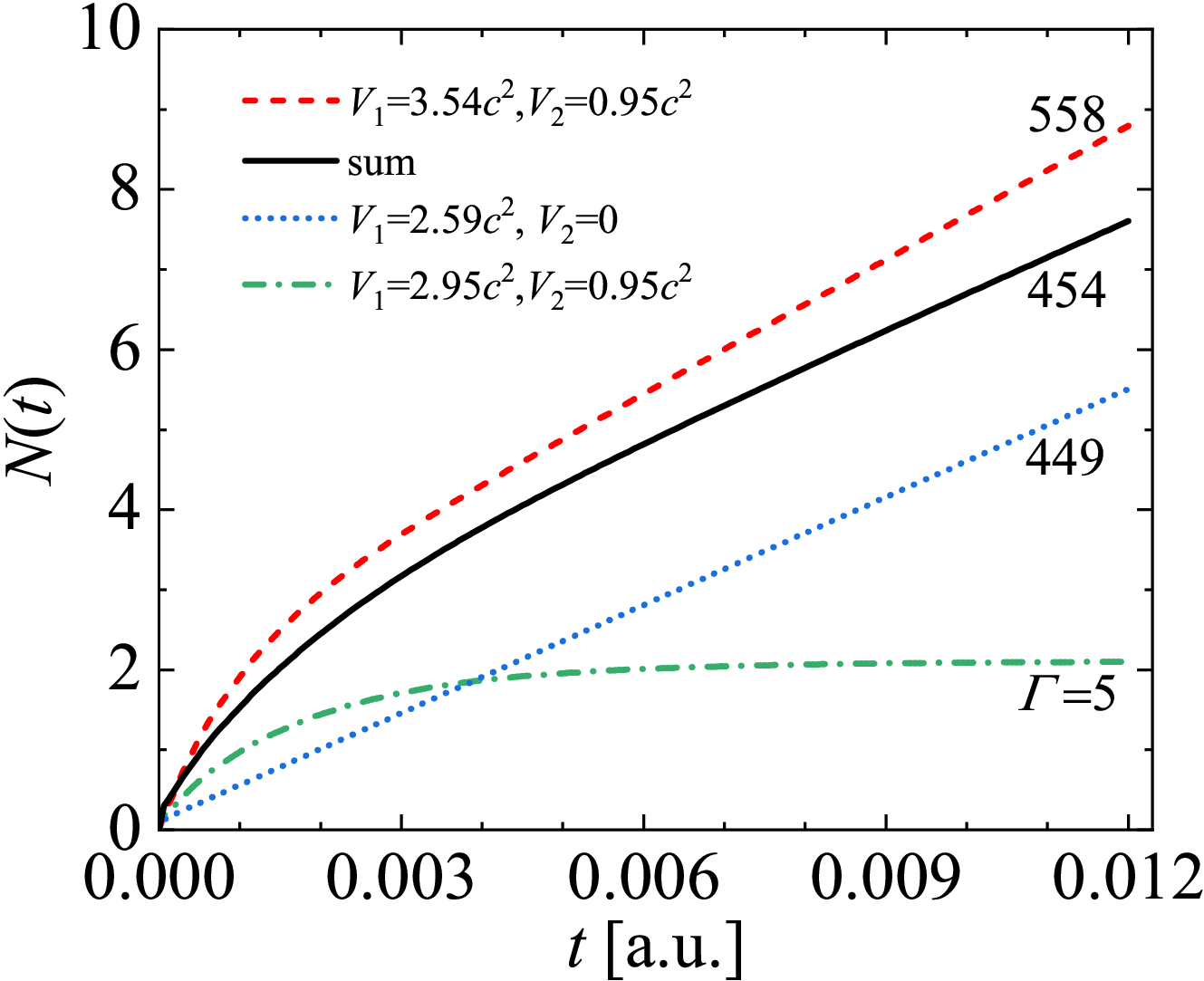}%
\caption{Time evolution of the number of positrons created in the asymmetric potential well with $V_1=3.54c^2, V_2=0.95c^2$ (red dashed line), the asymmetric potential well with $V_1=2.95c^2, V_2=0.95c^2$ (green dash-dotted line), and the step potential with $V_1=2.95c^2, V_2=0$ (blue dotted line). The black solid line shows the sum of the last two cases. The slopes of these lines $\Gamma$ in the long-time limit are marked on each line. Other parameters are $w=0.3/c, d=0.2\,\mathrm{a.u.}, N_z=2048, L=8\,\mathrm{a.u.}$.}
\label{fig:Ehancement-Nt}
\end{figure}

Since the asymmetric potential well with $V_1 = 3.54c^2, V_2 = 0.95c^2$ can be viewed as a combination of a symmetric potential well with $V_1=V_2=0.95c^2$ and a step potential with $V_1= 2.59c^2, V_2=0$, we examine the relation between the growth rates of the number of positrons created in the asymmetric potential well, the symmetric potential well, and the step potential.
Considering that when a symmetric potential well is combined with a supercritical step potential to form a supercritical asymmetric potential well, the symmetric potential well will overlap with the negative energy continuum to produce particles.
Therefore, we use the asymmetric potential well with $V_1 = 2.95c^2, V_2 = 0.95c^2$ to characterize pair creation within the symmetric potential well.
According to Fig. \ref{fig:Ehancement-Nt-cb}(b), this replacement is approximately reasonable.
The time evolution of the number of positrons created in the asymmetric potential well with $V_1 = 3.54c^2, V_2 = 0.95c^2$, the asymmetric potential well with $V_1 = 2.95c^2, V_2 = 0.95c^2$,  and the step potential with $V_1= 2.59c^2, V_2=0$ is depicted in Fig. \ref{fig:Ehancement-Nt}.
The red dashed, green dash-dotted, and blue dotted lines correspond to the time evolution of the positron number for the asymmetric potential well with $V_1 = 3.54c^2, V_2 = 0.95c^2$, the asymmetric potential well with $V_1 = 2.95c^2, V_2 = 0.95c^2$, and the step potential, respectively.
The black solid line represents the sum of the positron number corresponding to the red dashed line and the green dash-dotted line.
As can be seen from the figure, in the long-time limit, the sum of the positron growth rates for the symmetric potential well ($5$) and the step potential ($449$) is less than that for the asymmetric potential well ($558$).
Consequently, pair creation can be enhanced by combining a symmetric potential well with a supercritical step potential.
The reason for this enhancement is that the presence of the potential well leads to scattering resonance (see Fig. \ref{fig:varying-v2-NE-pe}(a)) when electrons created by the continuum-continuum overlap cross the potential well, which enhances pair creation.
In contrast, no such resonance occurs in a step potential, where only the continuum-continuum overlap exists.
Thus, it is the combined effect of the symmetric potential well and the step potential that results in a higher growth rate of created particles than the sum of the individual growth rates for these two potentials.
Moreover, in the long-time limit, the number of particles created by the bound-continuum overlap tends to a constant.
Only the continuum-continuum overlap region continues to create particles at a constant rate.
Hence, the enhancement effect mainly comes from the overlap region between the positive and negative energy continuum.
This discovery provides us with a new way to enhance pair creation.

\section{Conclusion and Outlook}
\label{sec:four}

In summary, we investigated electron-positron pair creation in a supercritical static asymmetric potential well composed of a subcritical potential and a supercritical potential using computational quantum field theory.
To explain the discrete peaks in the positron energy spectrum, an analytical formula for the bound state energy levels in a subcritical asymmetric potential well is derived and extended to the supercritical asymmetric potential well in two ways.
One is to fit the functional relationship between the bound state energy and a larger potential height according to the analytical formula, and then use the relationship to the supercritical asymmetric potential well.
The other is to directly extent the analytical formula to the complex energy region.
Employing these two methods, the bound state energy levels in the supercritical asymmetric potential well are computed and compared with those obtained from CQFT.
The result shows that both methods can accurately predict the positions of the bound states.
The second one can not only determine the positions of the bound states (the real part of the solutions to the analytical formula), but also predict the pair creation rate (two times the imaginary part of the solutions to the analytical formula).
Based on the above outcomes, we confirmed that the discrete peaks in positron energy momentum are indeed induced by the formation of resonate states due to the overlap between the bound states and the negative energy continuum.
From the electron energy spectrum, we found that adjusting the subcritical potential height allows for creating energy-concentrated electrons.
Since adjusting the parameters of supercritical fields is challenging, this approach may be more favorable for obtaining energy-concentrated electron beams in experiments.

Moreover, we also explored the time evolution of pair creation in the supercritical static asymmetric potential well.
It is found that in the initial stage, both the bound-continuum and continuum-continuum overlaps contribute to pair creation.
The production process is similar to that in the absence of the subcritical potential.
However, once the bound state energy levels are fully occupied by created electrons, the Pauli exclusion principle dictates that only the continuum-continuum overlap region continues to produce electrons at a constant rate.
Furthermore, we found that the coexistence of the bound-continuum and continuum-continuum overlaps can enhance pair creation, and the enhancement primarily localized in the continuum-continuum overlap region.
This finding provides a new way for optimizing field configurations to maximize particle yield and a promise for experimental verification of vacuum pair production.

In this study, we observed that the positron energy spectrum contains many small peaks between the discrete peaks.
The origin of these small peaks, such as whether they are formed by interference between bound states, requires further investigation.
For supercritical symmetric potential wells, the positions of bound states dived in the negative energy sea can be determined by calculating the maxima of the transmission coefficient formula. However, there is no such transmission coefficient formula for the supercritical asymmetric potential well.
Therefore, whether an analytical method can be developed to determine the positions of bound states in an supercritical asymmetric potential well is an open question.
Finally, using complex scaling method to calculate the eigenvalues of the Hamiltonian to determine bound state energy levels in an supercritical asymmetric potential well is also worth further research.

\begin{acknowledgments}
The work is supported by the National Natural Science Foundation of China (NSFC) under Grants No. 11974419 and No. 11705278, the Strategic Priority Research Program of Chinese Academy of Sciences (Grant No. XDA25051000, XDA25010100), and the Fundamental Research Funds for the Central Universities (No. 2023ZKPYL02, 2025 Basic Sciences Initiative in Mathematics and Physics).
\end{acknowledgments}

\appendix
\section{Calculation of the energy and width of resonant states in a supercritical symmetric potential well}
\label{appa}

For a symmetric potential well, such as the potential (\ref{eqn:Potential}) with $V_1=V_2$, when its depth is smaller than $-2c^2$, the bound states will dive into the Dirac sea and form resonant states. The resonant state energy levels can be extracted from the maxima of the transmission coefficient \cite{Greiner4}:
\begin{equation}
T=\frac{1}{1+\left(\frac{1-\gamma^{2}}{2\gamma}\right)^{2}\sin^{2}
\left(p_{2}d\right)},
\end{equation}
where $\gamma$, $p_2$, and $d$ are given in Sec. \ref{sec:sub3a}.
According to this formula, the resonant state energy is computed and shown in the third column of Table \ref{tab:comparison2}.
\begin{table*}[!ht]
\renewcommand{\arraystretch}{1.5}
\setlength{\tabcolsep}{5pt}
\begin{center}
   \caption{\label{tab:comparison2} Comparison of the energy and width of the resonant states computed by the bound state energy level formula and by the transmission coefficient. $E_\mathrm{boud}$ and $E_\mathrm{tran}$ represent the resonant state energy obtained using the above two methods, respectively. $\Delta E_\mathrm{FWHM}$ denotes the full width at half maximum (FWHM) of the resonant states as determined by the transmission coefficient.
   The field parameters are $V_1=3.0c^2,\,V_2=3.0c^2,\,w\rightarrow0$, and $d=10/c$.}
   \begin{tabular}{cccccc} \hline \hline
    \makecell[c]{Resonant state \\number}  & $ E_\mathrm{boud}\,\,[\mathrm{c^2}]$ & $ E_\mathrm{tran}\,\,[\mathrm{c^2}]$ & $\Delta E_\mathrm{FWHM}/2\,\,[\mathrm{c^2}]$ & $\big|\frac{\Re(E_\mathrm{boud})-E_\mathrm{tran}}{E_\mathrm{tran}}\big|\,\,[\%]$ & $\frac{|\Im(E_\mathrm{boud})-\Delta E_\mathrm{FWHM}/2|}{\Delta E_\mathrm{FWHM}/2}\,\,[\%]$ \\ \hline
    $1$ & $-1.136261+0.023681\mathrm{i}$ & $-1.137904$ & $0.024325$ & $0.1444$ & $2.6497$ \\ 
    $2$ & $-1.393799+0.031086\mathrm{i}$ & $-1.394031$ & $0.032695$ & $0.0167$ & $4.9215$ \\ 
    $3$ & $-1.626543+0.026957\mathrm{i}$ & $-1.625859$ & $0.028391$ & $0.0421$ & $5.0502$ \\
    $4$ & $-1.819881+0.016635\mathrm{i}$ & $-1.818990$ & $0.017261$ & $0.0490$ & $3.6290$ \\
    $5$ & $-1.952213+0.005209\mathrm{i}$ & $-1.951813$ & $0.005316$ & $0.0205$ & $1.9995$ \\
     \hline \hline
   \end{tabular}
   \end{center}
\end{table*}
The half width at half maximum of the resonant states determined by the transmission coefficient is shown in the forth column of Table \ref{tab:comparison2}.
The energy of the resonant states computed by the bound state energy level formula (\ref{eqn:bsel1}) is shown in the second column of Table \ref{tab:comparison2}.
We can see that when the bound states dive into the negative energy continuum and form resonant states, the formula (\ref{eqn:bsel1}) has the complex energy solutions.
Do the complex energy solutions make sense? The answer is yes.
By comparing the real and imaginary part of $E_\mathrm{boud}$ with $E_\mathrm{tran}$ and $\Delta E_\mathrm{FWHM}/2$, respectively, we find that the maximum relative error for the former (the fifth column of Table \ref{tab:comparison2}) is less than $0.15\%$ and for the latter (the sixth column of Table \ref{tab:comparison2}) is about $5\%$.
This indicates that the real part of $E_\mathrm{boud}$ can be used to estimate the energy of resonant states, while its imaginary part can predict the width of resonant states.

\end{document}